\documentclass[twocolumn,showpacs,floats,floatfix,superscriptaddress,aps,pra]{revtex4-1}
\usepackage{amsfonts}
\usepackage{amssymb}
\usepackage{amsmath}
\usepackage{calc}
\usepackage{graphicx}
\usepackage{bm}

\usepackage[normalem]{ulem}
\usepackage{amsmath,amssymb}
\usepackage{multirow}
\usepackage{xcolor,soul}
\usepackage{srcltx}
\usepackage{hyperref,graphicx}

\def\be{ \begin{equation}}
\def\ee{ \end{equation}}
\def\bea{ \begin{eqnarray}}
\def\eea{ \end{eqnarray}}
\def\bse{ \begin{subequations}}
\def\ese{ \end{subequations}}
\def\bc{ \begin{center}}
\def\ec{ \end{center}}

\begin{document}

\author{Giuseppe Della Valle}
\affiliation{Dipartimento di Fisica, Politecnico di Milano and Istituto di Fotonica e Nanotecnologie del Consiglio Nazionale delle Ricerche, Piazza L. da Vinci 32, I-20133 Milano, Italy}
\author{Stefano Longhi}
\affiliation{Dipartimento di Fisica, Politecnico di Milano and Istituto di Fotonica e Nanotecnologie del Consiglio Nazionale delle Ricerche, Piazza L. da Vinci 32, I-20133 Milano, Italy}

\title{Spectral and transport properties of time-periodic \\$\mathcal{PT}$-symmetric tight-binding lattices}
\date{\today }

\begin{abstract}
We investigate the spectral properties and dynamical features of a time-periodic $\mathcal{PT}$-symmetric Hamiltonian on a one-dimensional tight-binding lattice. It is shown that a high-frequency modulation can drive the system under a transition between the broken-$\mathcal{PT}$ and the unbroken-$\mathcal{PT}$ phases. The time-periodic modulation in the unbroken-$\mathcal{PT}$ regime results in a significant broadening of the quasi-energy spectrum, leading to a {\it hyper-ballistic} transport regime. Also, near the $\mathcal{PT}$-symmetry breaking the dispersion curve of the lattice band becomes linear, with a strong reduction of quantum wave packet spreading.
\end{abstract}

\pacs{
03.65.-w, 
11.30.Er, 
72.10.Bg, 
42.82.Et, 
}
\maketitle

\section{Introduction}

Non-Hermitian Hamiltonians play an important role in physics \cite{Moiseyev}. They are introduced, for example, to model open systems and dissipative phenomena in a wide variety of physical contexts  (see, for instance, \cite{Moiseyev,Rotter}). A special class of non-Hermitian Hamiltonians is provided by complex potentials having parity-time ($\mathcal{PT}$) symmetry \cite{Buslaev_JPA_1993, Bender_PRL_98, Bender_RPP_2007}, that is invariance under simultaneous parity transform ($\mathcal{P}$: $\hat{p} \rightarrow -\hat{p}$, $\hat{x} \rightarrow -\hat{x}$, where $\hat{p}$ and $\hat{x}$ stand for momentum and position operators, respectively) and time reversal ($\mathcal{T}$: $\hat{p} \rightarrow -\hat{p}$, $\hat{x} \rightarrow \hat{x}$, $i\rightarrow-i$). An important property of $\mathcal{PT}$ Hamiltonians is to admit of an entirely real-valued energy spectrum below a phase transition symmetry-breaking point. Unbroken $\mathcal{PT}$-symmetric Hamiltonians possess a complete set of eigenvectors and have originally attracted great attention for the possibility to formulate  a consistent quantum mechanical theory in a non-Hermitian framework \cite{Bender_RPP_2007,Bender_PRL_98,Mostafazadeh_JPA_02,Note0}. The appeal of such an idea originated in the years-long emphasized
relevance of systems with $\mathcal{PT}$-symmetry in quantum field theory (see e.g. \cite{Bender_PRL_04, Shalaby_PRD_09}), but have found interest and applications in several other physical fields, such as in classical mechanics \cite{Bender_JMP_1999}, magnetohydrodynamics \cite{Guenther_JMP_2005}, cavity quantum electrodynamics \cite{Plenio_RMP_1998}, and electronics \cite{Kottos}. Very recently, the peculiar features of $\mathcal{PT}$-symmetry have found important applications in optics, with the theoretical proposal and experimental demonstration of innovative materials and optical devices based on $\mathcal{PT}$-invariant complex dielectric functions \cite{Muga,El-Ganainy_OL_07, Makris_PRL_08, Klaiman_PRL_08, Mostafazadeh_PRL_09, Longhi_PRL_09, Guo09,Ruter_NP_10, Feng2011,Kivshar12,Regensburger_Nature_12,Feng12}. $\mathcal{PT}$ optical media offer rather  unique transport properties, including, as example, double refraction, power oscillations,  and nonreciprocal diffraction patterns \cite{Makris_PRL_08}, unidirectional Bragg scattering and invisibility \cite{Longhi_PRL_09, Lin_PRL_2011, Regensburger_Nature_12, LonghiPRA10, Longhi_JPA_2011,Graefe11,Feng12}, giant Goos-H\"anchen shift \cite{Longhi_PRA_2011}, and simultaneous perfect absorption and laser behaviour \cite{Longhi_PRA_2010, Chong_PRL_2011}. 

Though the physics of $\mathcal{PT}$-symmetric Hamiltonians has been widely explored, most of the attention has been devoted to static (i.e.~time-independent) potentials. It is only very recently that the study of time-dependent $\mathcal{PT}$-symmetric potentials has attracted an increasing attention \cite{Faria_JPA_2006, Faria_LasPhys_2007, Kottoschaos,Wu_PRA_2012, Longhi_PRB_09,Mo1,Mo2,Gong}. The time-dependent part of the Hamiltonian considered in most of such previous studies is an Hermitian driving term acting on a static non-Hermitian potential, whereas a genuine (and eventually strong) non-Hermitian driving has been investigated so far solely for a simple low-dimensional (dimeric) $\mathcal{PT}$-symmetric system \cite{Mo1,Mo2} and for the $\mathcal{PT}$-symmetric kicked rotor model \cite{Kottoschaos} in connection to dynamical localization and quantum chaos. On the other hand, time-dependent Hermitian Hamiltonians are of fundamental importance in many branches of physics, with particular interest to time-periodic Hamiltonians entailing a plethora of phenomena ranging from Rabi oscillations (see e.g. \cite{Scully}) and Autler-Townes splitting \cite{AT_PR_55}, to dynamic localization \cite{Dunlap_PRB_86} and coherent destruction of tunneling \cite{Grossmann_PRL_91, Della_Valle_PRL_2006}, just to mention a few (see also \cite{Grifoni_PR_98} for a review on driven quantum tunneling). Therefore, it  is of great interest to extend the theoretical investigations of coherently-driven quantum systems within the context of $\mathcal{PT}$-symmetric Hamiltonians, especially for spatially-extended systems, like in complex crystals, where $\mathcal{PT}$ symmetry can strongly affect the transport properties.\\

In the present paper we consider a time-dependent $\mathcal{PT}$-symmetric potential defined on a one-dimensional tight-binding lattice with nearest neighbor hopping rate, in which a time-periodic complex modulation of the potential acts as a non-Hermitian $\mathcal{PT}$ periodic driving of a static Hermitian lattice. We show that in the time-periodic system the frequency and amplitude of the driving can control the $\mathcal{PT}$ symmetry phase transition, and provide a detailed investigation of the spectral properties and dynamical features of the driven lattice below the  $\mathcal{PT}$ symmetry breaking point. A comparison with the un-driven Hamiltonian and with Hermitian periodic driving (that is a real potential modulation) is given to highlight the peculiar effects of non-Hermitian driving on the quasi-energy spectrum of the system and related transport properties. In particular, we show that the time-periodic modulation in the unbroken $\mathcal{PT}$ regime results in a significant broadening of the quasi-energy spectrum, leading to a {\it hyper-ballistic} transport regime and  to a  strong reduction of quantum wave packet spreading near the $\mathcal{PT}$-symmetry breaking threshold.\\The paper is organized as follows. In Sec. II we introduce the time-periodic $\mathcal{PT}$-symmetric lattice model and investigate the effects of the time-periodic modulation on the $\mathcal{PT}$-symmetry breaking point of the system. In Sec. III we discuss the peculiar spectral features and transport properties of the lattice under strong high frequency modulation, showing the appearance of enhanced ballistic transport and reduction of wave packet dispersion as the $\mathcal{PT}$-symmetry breaking boundary is approached. Finally, in Sec.~IV the main conclusions and future developments are outlined.

\section{A time-periodic $\mathcal{PT}$-symmetric tight-binding lattice}
We consider here a $ \mathcal{PT}$-invariant tight-binding Hamiltonian describing the hopping dynamics of a single particle on a one-dimensional binary  superlattice, with a superposed modulation of the  'imaginary' part of the site energies (loss or gain term) with amplitudes $ \pm \Delta(t)$ at alternating lattice sites. The Hamiltonian of the superlattice reads:
\begin{equation}\label{Acca}
H(x,t) = H_0(x) + H_{drive}(x,t) 
\end{equation}
\noindent where
\begin{equation}\label{Acca0}
H_0(x) = - \hbar \kappa \sum_{n=-\infty}^{\infty} \left(  |n \rangle \langle n+1| + | n+1 \rangle \langle n | \right), 
\end{equation}
\begin{equation}\label{AccaD}
H_{drive}(x,t) = i \hbar \Delta(t) \sum_{n=-\infty}^{\infty}  (-1)^n | n \rangle \langle n |
\end{equation}
In above Eqs.~(\ref{Acca0})-(\ref{AccaD}), $\kappa$ is the hopping rate between adjacent lattice sites and $|n \rangle$ is the Wannier state localized at the $n$-th site.  Note that the resulting total Hamiltonian $H$ is $\mathcal{PT}$-invariant, regardless the temporal dependence detailed by $\Delta(t)$. For a static (i.e. time-independent) real-valued $\Delta$ the $\mathcal{PT}$-invariant lattice model Eqs.(1-3) was previously investigated in Refs.\cite{Longhi_PRL_09,Muss} and was shown to be always in the $\mathcal{PT}$ broken phase for any infinitesimally-small value of $\Delta$. For a time-periodic but imaginary $\Delta(t)$, corresponding to Hermitian ac-driving, the lattice model Eqs.(1-3) was previously investigated in Refs.\cite{Sta1,Sta2} and it was shown to realize, under certain modulation conditions, approximate suppression of particle spreading (dynamic localization).\\
The state vector of the system $|\psi(t) \rangle$ in Wannier basis representation can be written as:
\begin{equation}
|\psi(t) \rangle = \sum_n c_n(t) | n \rangle,
\end{equation}
\noindent where $c_n(t)$ is the complex amplitude for occupation of the $| n \rangle$ Wannier state. Given above decomposition of the state vector, standard projection technique provides the following evolution equations for the amplitudes $c_n(t)$:
\begin{equation}\label{CMEs}
i \dot{c_n} = - \kappa \left(c_{n-1} + c_{n+1} \right) + i (-1)^n \Delta(t)c_n
\end{equation}
where the dot stands for the derivative with respect to time $t$.
Let us assume that the driving amplitude $\Delta(t)$ is a time-periodic function of period $T$ with zero mean value, i.e. $\int_0^T dt \Delta(t)=0$. The spectral and transport properties of the ac-driven lattice can be throughly investigated by computing the quasi-energy spectrum $\mathcal{E}(q)$ of $H$, where the quasi-momentum $q$ varies in the interval $-\pi \leqslant q < \pi$ (see e.g.~\cite{Grifoni_PR_98} and references therein). In the following analysis, we will consider a square-wave ac modulation $\Delta(t)$ for the sake of definiteness, however similar results are obtained for different types of modulation, such as for a sinusoidal modulation. For a square-wave driving $\Delta(t) = \Delta_0\; {\rm square}(\omega t + \theta)$, where $\omega = 2\pi/T$ is the modulation frequency and $\theta$ an initial phase. In this case the monodromy matrix ${\bf M}$ (connecting the solution to the time-periodic system of Eq.~(\ref{CMEs}) in Fourier space over one oscillation cycle) can be analytically determined as ${\bf M} = {\bf M}_2 {\bf M}_1$, with:
\begin{subequations}\label{M1M2}
\begin{align}
{\bf M}_1 = \left[\begin{array}{cc}
C+(\Delta_0/\lambda)S & i(\rho/\lambda)S\\
i(\rho^*/\lambda)S & C-(\Delta_0/\lambda)S \end{array}\right]\\
{\bf M}_2 = \left[\begin{array}{cc}
C-(\Delta_0/\lambda)S & i(\rho/\lambda)S\\
i(\rho^*/\lambda)S & C+(\Delta_0/\lambda)S \end{array}\right]\\ \nonumber
\end{align}
\end{subequations}
\noindent where $C = \cosh (\lambda T/2)$, $S = \sinh(\lambda T/2)$, $\rho = \kappa (1+{\rm e}^{iq})$ and $\lambda = \sqrt{\Delta_0^2-|\rho|^2}$. Details are given in the Appendix.\\
\begin{figure}[t]
\includegraphics[width=8cm]{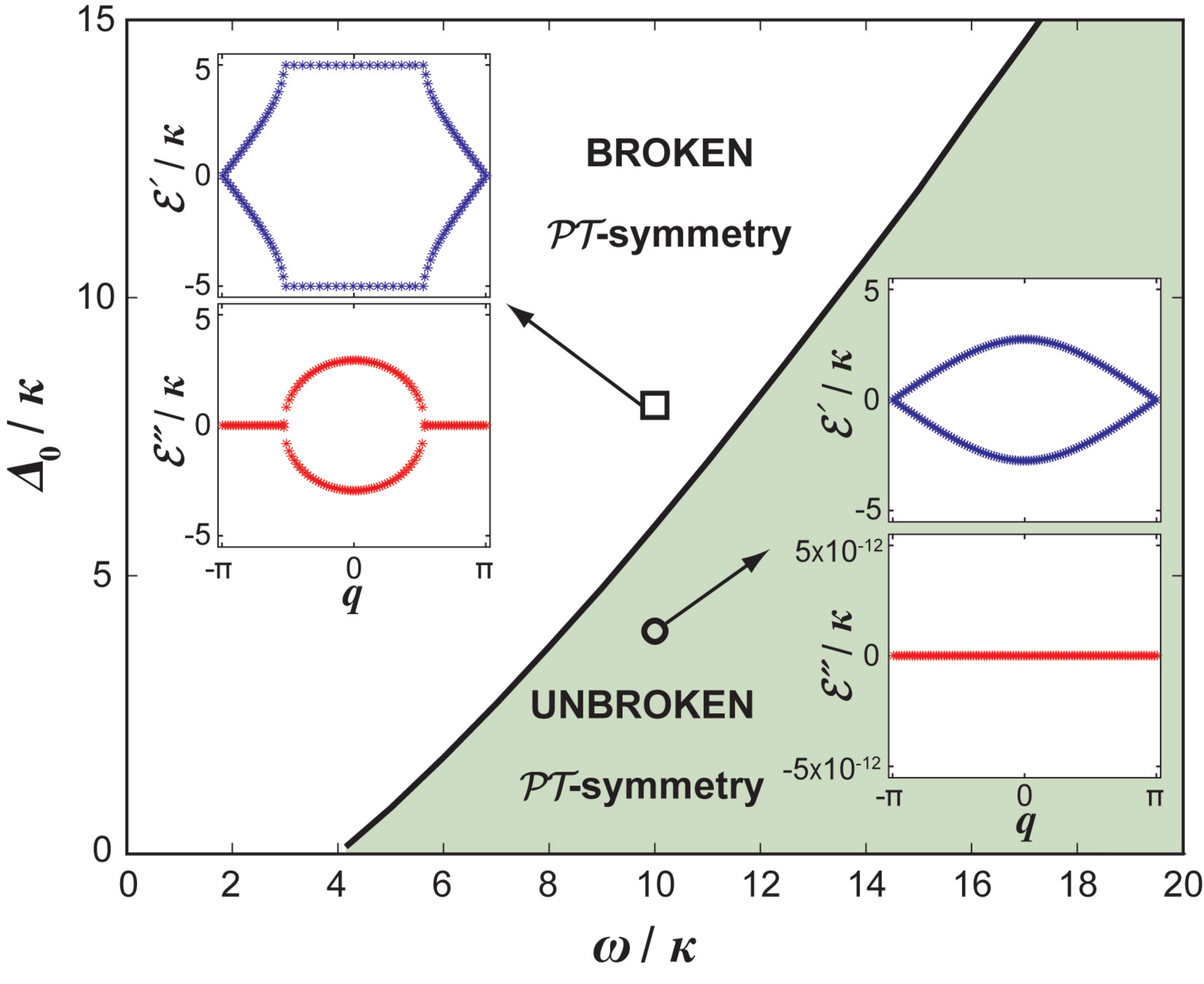}
\caption{(Color online). Phase diagram of the Hamiltonian Eqs.~(\ref{Acca})-(\ref{AccaD}) with square wave time-periodic complex potential of frequency $\omega$ and amplitude $\Delta_0$. Inset shows real part (upper panel) and imaginary part (lower panel) of the quasi-energies for two points at $\omega = 10 \kappa$, one above (left panels) and one below (right panels) $\mathcal{PT}$-symmetry breaking (solid curve).}
\label{transition}
\end{figure}
Once the complex eigenvalues $\eta(q)$ of ${\bf M}$ are computed as a function of the quasi-momentum $q$, the real part $\mathcal{E}'$ and imaginary part $\mathcal{E}''$ of the quasi-energy spectrum can be evaluated as follows (see the Appendix for details):
\begin{subequations}\label{QEs}
\begin{align}
& \mathcal{E}'(q) = \frac{1}{T}\arg[\eta(q)],\\
& \mathcal{E}''(q) = -\frac{1}{T}\log[|\eta(q)|],
\end{align}
\end{subequations}

\begin{figure}[b]
\includegraphics[width=8.5cm]{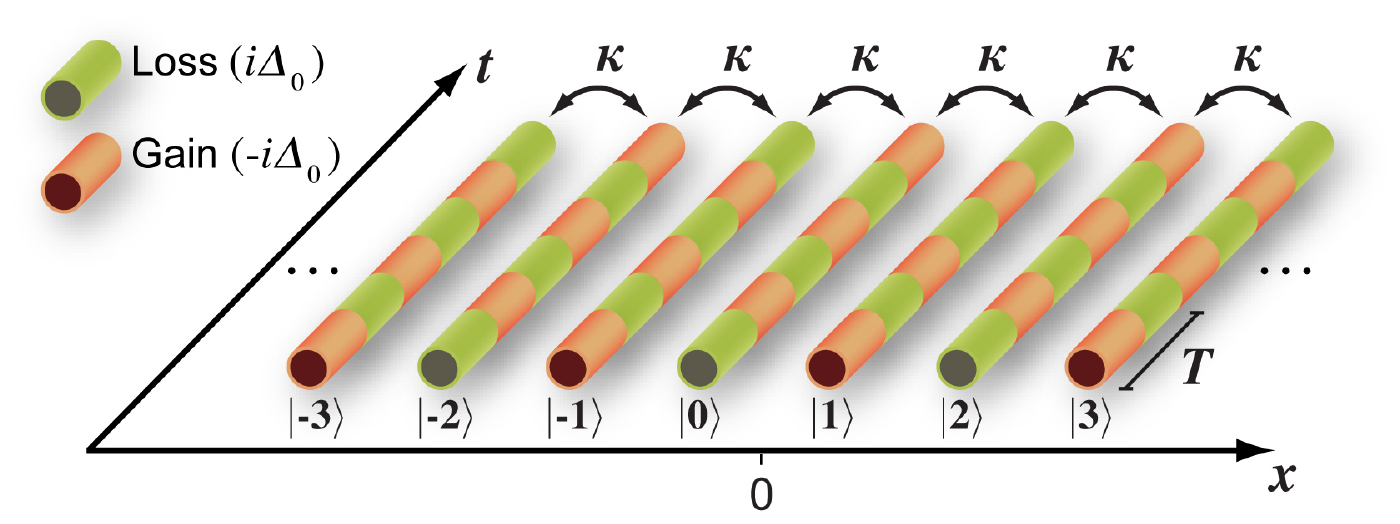}
\caption{(Color online). Schematic of a tunneling-coupled optical waveguide array realizing the $\mathcal{PT}$-symmetric Hamiltonian of Eqs.~(\ref{Acca})-(\ref{AccaD}). Here time-periodic modulation is replaced by periodic modulation along the optical axis $t$ of alternating  optical loss and gain sections.}
\label{implementatio}
\end{figure}

The nature of the $\mathcal{PT}$-symmetry in the $(\omega,\Delta_0)$ plane of parameters can be throughly determined by inspecting $\mathcal{E}''$ (Fig.~1). We find that the in the low-frequency regime, the system exhibits a broken $\mathcal{PT}$-symmetry phase, with complex quasi-energies [c.f. left inset in Fig.~1]. Above a minimum value of the frequency $\omega_m \simeq 4 \kappa$, the system can be driven to the unbroken $\mathcal{PT}$-symmetric phase (shaded area), having a purely real quasi-energy spectrum [c.f. right inset in Fig.~1], provided that the driving amplitude $\Delta_0$ is lower than a threshold value which is an almost linear function of the modulation frequency [solid line in Fig.~1]. Therefore, contrary to the static case \cite{Longhi_PRL_09,Muss}, a broad range of modulation parameters exists where the driven Hamiltonian defined by Eqs.~(\ref{Acca}-\ref{AccaD}) has a real-valued energy spectrum and can exhibit potentially interesting novel transport properties.\\

It is worth mentioning briefly that the time-periodic $\mathcal{PT}$-symmetric Hamiltonian above discussed can find a physical implementation as example in optical systems. Optical realizations of $\mathcal{PT}$-symmetric photonic lattices, based on light propagation in coupled waveguides with alternating regions of optical gain and absorption, have been discussed in many previous works (see, e.g.,  ~\cite{El-Ganainy_OL_07, Makris_PRL_08, Klaiman_PRL_08, Longhi_PRL_09}). In Fig.~2 we show a sketch of a possible implementation of the one-dimensional tight-binding lattice with $\mathcal{PT}$ symmetric time-periodic driving, in the form of a linear array of tunneling-coupled optical waveguides with an alternation of loss and gain sections along propagation axis $t$. This represents somehow a generalization of the original approach to $\mathcal{PT}$ symmetric photonic lattices, recently demonstrated experimentally for a static $\mathcal{PT}$-symmetric directional coupler~\cite{Ruter_NP_10}. Another possibility to implement our Hamiltonian is to exploit the recent proposal of mimicking $\mathcal{PT}$-symmetric optical crystals with a network of coupled fiber loops ~\cite{Regensburger_Nature_12}. This would represent a time-discrete version of the photonic lattice of Fig.~2, where time evolution is implemented as a repeated circulation in the fiber loops (instead of space propagation along the optical axis of a photonic structure), and the loops can provide on demand gain or loss in equal amount by means of optical amplifiers and amplitude modulators.

\section{Spectral and transport properties}

The quasi-energy spectrum in a time-periodic lattice is known to govern the transport properties of one particle in the lattice. It is worth noting that ac-driving is a powerful and ubiquitous technique of quantum control within Hermitian systems, allowing one to shrink the quasi-energy spectrum and eventually collapse the quasi-energies, giving rise to dramatic effects on transport, like coherent suppression of tunneling and dynamic localization (see e.g.~\cite{Grifoni_PR_98} and references therein). In this section we investigate  the transport properties of non-Hermtian ac-driving of the $\mathcal{PT}$-symmetric lattice, highlighting the appearance of a novel transport regime, referred to as {\it hyper-ballistic} motion, and the possibility to strongly reduce wave packet spreading.
\begin{figure}[t]
\includegraphics[width=8cm]{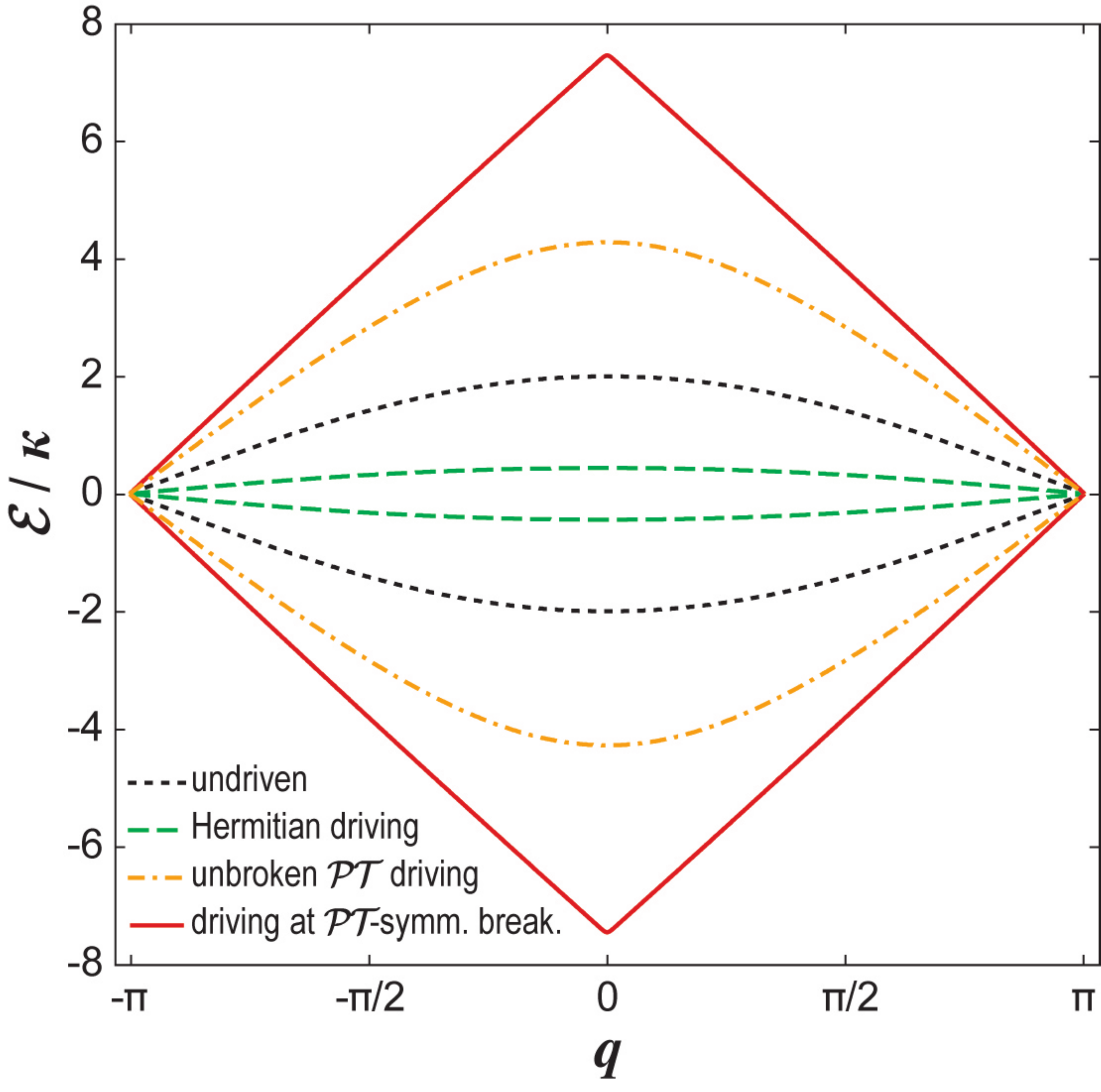}
\caption{(Color online). Quasi-energy spectra at frequency modulation $\omega = 15 \kappa$ for different values of $\Delta_0$: 0 (dotted); $-i 11.998\kappa$ (dashed); $10\kappa$ (dash-dotted); $11.998\kappa$ (solid), corresponding to $\mathcal{PT}$-symmetry breaking.}
\label{QEexpansion}
\end{figure}

In Fig.~3 we compare the quasi-energy spectra of the Hamiltonian considered in the previous section for a fixed value of the modulation frequency $\omega = 15 \kappa$ and different values of the driving amplitude $\Delta_0$. The energy spectrum of the undriven Hamiltonian $H_0$ is also shown (dotted line). We found that as $\Delta_0$ is increased, the quasi-energy spectrum is progressively broadened, contrary to the case of Hermitian driving (also reported in dashed line for comparison) that always gives rise to a shrink of the spectrum \cite{Sta1}. The non-Hermitian nature of the periodic driving (even in the unbroken $\mathcal{PT}$-symmetric phase) is thus qualitatively distinct from an Hermitian driving acting on the same static Hamiltonian. Moreover, as the $\mathcal{PT}$-symmetry breaking is approached, the quasi-energy spectrum attains a progressive linear dispersion relation, culminating in a zero group-velocity dispersion $d^2\mathcal{E}/dq^2$ in the whole Brillouin zone, with the exception of the zone center and edges.

The impact of the spectral features above detailed on the transport properties of the system is twofold: (i) the broadening of the quasi-energies is expected to enhance the ballistic motion of the particle, i.e.~to introduce a  novel regime (with no counterpart in Hermtian lattices) that we can refer to as {\it hyper-ballistic} motion: the group velocity in the lattice of a particle wave packet is not limited by the hopping rate between adjacent sites; (ii) the linearization of the dispersion curve near the $\mathcal{PT}$ symmetry breaking point results in a strong reduction of the spreading (diffraction) for a particle wavepacket.\\

\begin{figure}[b]
\includegraphics[width=8cm]{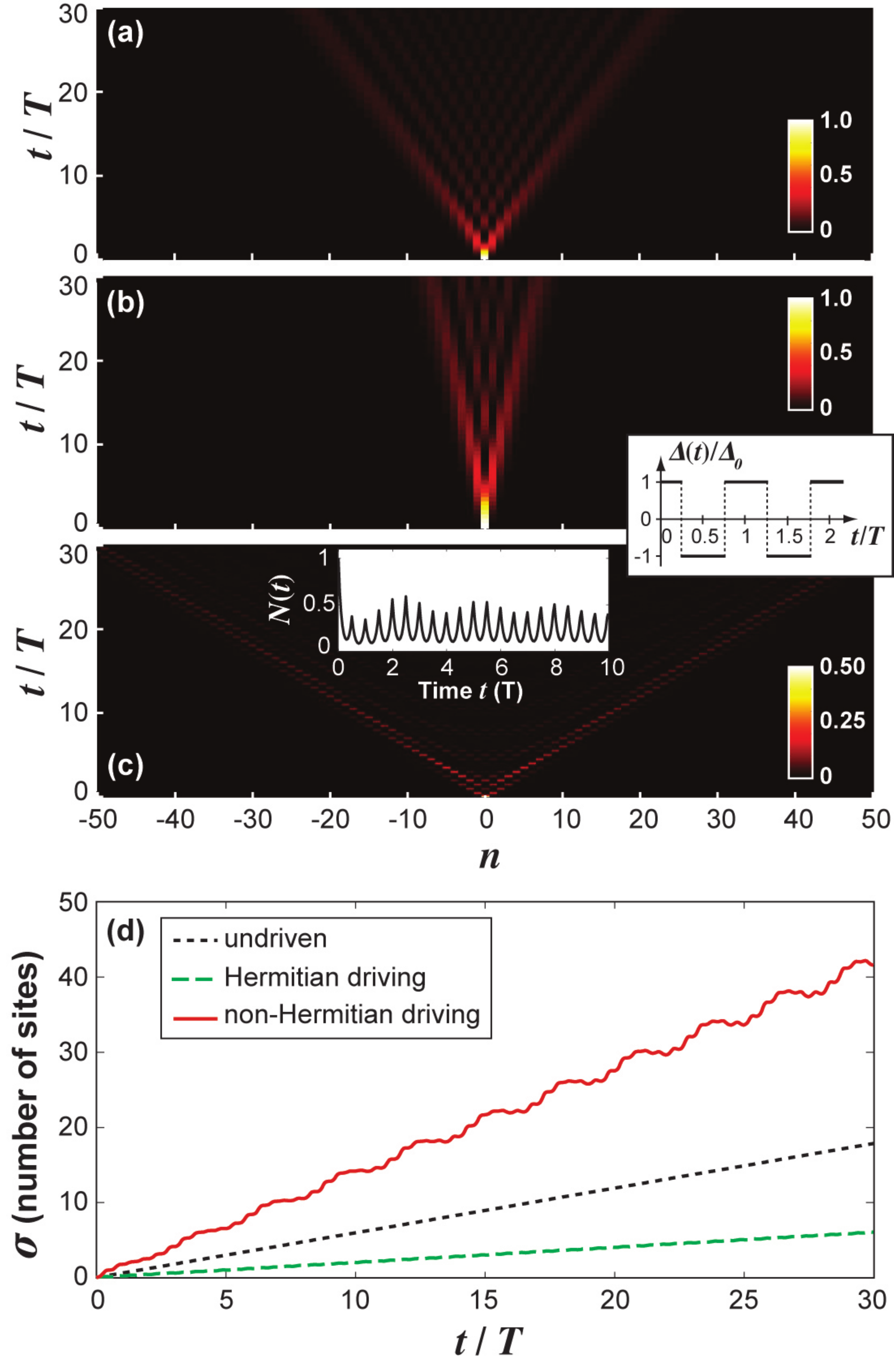}
\caption{(Color online). Occupation probability $|c_n(t)|^2$ for initial condition $c_n(0) = \delta_{n,0}$ and modulation frequency $\omega = 15 \kappa$ at different modulation amplitudes: (a) $\Delta_0 = 0$ (undriven system); (b) $\Delta_0 = -i 10.7\kappa$ (Hermitian driving); (c) $\Delta_0 = 10.7 \kappa$ (non-Hermitian driving close to the $\mathcal{PT}$-symmetry breaking). Inset in (b)-(c) shows the time-periodic driving considered in the simulations. Inset in (c) shows the norm of the vector state for the first 10 periods of the modulation under non-Hermitian driving. (d) Wave-packet variance $\sigma(t)$ corresponding to the three cases detailed in (a)-(c).}
\label{superdiff}
\end{figure}

We have checked the above predictions by numerical simulations of Eqs.~(\ref{CMEs}) for either narrow (single-site) and broad wave packet excitation. As is well known in the Hermitian ac-driving case, the phase (i.e. time origin) of the modulation $\Delta(t)$ has a noticeable impact onto the wave packet dynamics (see, for instance, \cite{Creffield_PRL_08}). For example, in the high-frequency limit, the intial phase $\theta$ is responsible to an additional phase term in the renormalized tunneling rate, as discussed in \cite{Creffield_PRL_08}. In the following analysis we will limit to consider a modulation function $\Delta(t)$ satisfying the constraint $\Phi(t-T/2) = - \Phi(t)$, where $\Phi(t) = \int_0^t dt' \Delta(t')$. This condition is satisfied, for instance, for a sinusoidal or square-wave modulation, under an appropriate choice of $\theta$. In this case, for the Hermitian ac-driving problem the renormalization of the tunneling rate affects the amplitude (but not the phase) of the effective hopping rate \cite{Creffield_PRL_08}.

The hyper-ballistic behavior of the system when driven close to the $\mathcal{PT}$-symmetry breaking can be investigated by solving Eqs.~(\ref{CMEs}) under initial excitation of an individual site (e.g. the $n=0$ site). In  Fig.~4 we show, as an example, the numerically-computed evolution of $|c_n(t)|^2$ obtained in a lattice with $201$ sites for the static Hamiltonian [panel (a)], for the dynamic Hamiltonian under driving at frequency $\omega = 15\kappa$ and amplitude $\Delta_0 = -i 10.7\kappa$ [panel (b)] corresponding to an Hermitian driving, and for driving at the same frequency and absolute value of the amplitude but in the non-Hermitian case, i.e. $\Delta_0 = 10.7\kappa$ [panel (c)]. The square-wave modulation $\Delta (t)$ is shown in the inset of Fig.~4(b)-(c). In the non-Hermitian case, the norm of the vector state $N(t) = \langle \psi(t)| \psi(t) \rangle=\sum_ n |c_n(t)|^2$ is bounded and does not diverge in the unbroken $\mathcal{PT}$ phase, however it is not conserved, as shown in the inset of Fig.~4(c)~\cite{Note1}.\\
Most interestingly, the non-Hermitian driving results in a strong enhancement of the ballistic spreading in the lattice which can be quantitatively estimated by computing the variance of the probability distribution as a function of time [Fig.~4(d)], defined as $\sigma(t) = \sqrt{\sum_n (n-\overline{n})^2 |c_n(t)|^2 / N(t) }$, where $\overline{n} = \sum_ n n|c_n(t)|^2 / N(t)$ is the average site occupation at time $t$.  It is seen that the ballistic spreading velocity $V = d\sigma / dt$ is about $2.4$ times larger as compared to the undriven Hamiltonian (limited to $V=2 \kappa$) for non-Hermitian driving close to the $\mathcal{PT}$-symmetry breaking [cf. solid and dotted lines in Fig.~4(d)]. This enhancement factor is in agreement with an analytical estimation of the effective hopping rate in the driven lattice based on asymptotic analysis of Eq.~(\ref{CMEs}). Actually, in the high frequency limit, the driven lattice described by Eq.~(\ref{CMEs}) turns out to be equivalent to a static effective lattice with renormalized hopping rate given by (see e.g. \cite{Grifoni_PR_98}) \cite{Note2}:
 \begin{equation}
 \kappa' = \kappa \frac{1}{T}\int_0^T e^{-2\Phi(t)} dt = \kappa \frac{1}{T}\int_0^T e^{2\Phi(t)} dt.
 \end{equation}
 For a square-wave modulation, one readily obtains $\kappa'  / \kappa = 2 \sinh(\Delta_0 T/2)/(\Delta_0 T)$, i.e. $\kappa' / \kappa \simeq 2.1$ for the parameters considered in our simulation. Therefore, the increase in the effective hopping rate corresponds to approximately the same increase in the ballistic rate. Note that Hermitian driving with similar parameters strongly reduces the effective hopping rate by a factor of $\sim4$ [cf. dotted and dashed lines in Fig.~4(d)].~Indeed, in the Hermitian driving the ratio $\kappa ' / \kappa$ is always smaller than one.\\

\begin{figure}[t]
\includegraphics[width= 8cm]{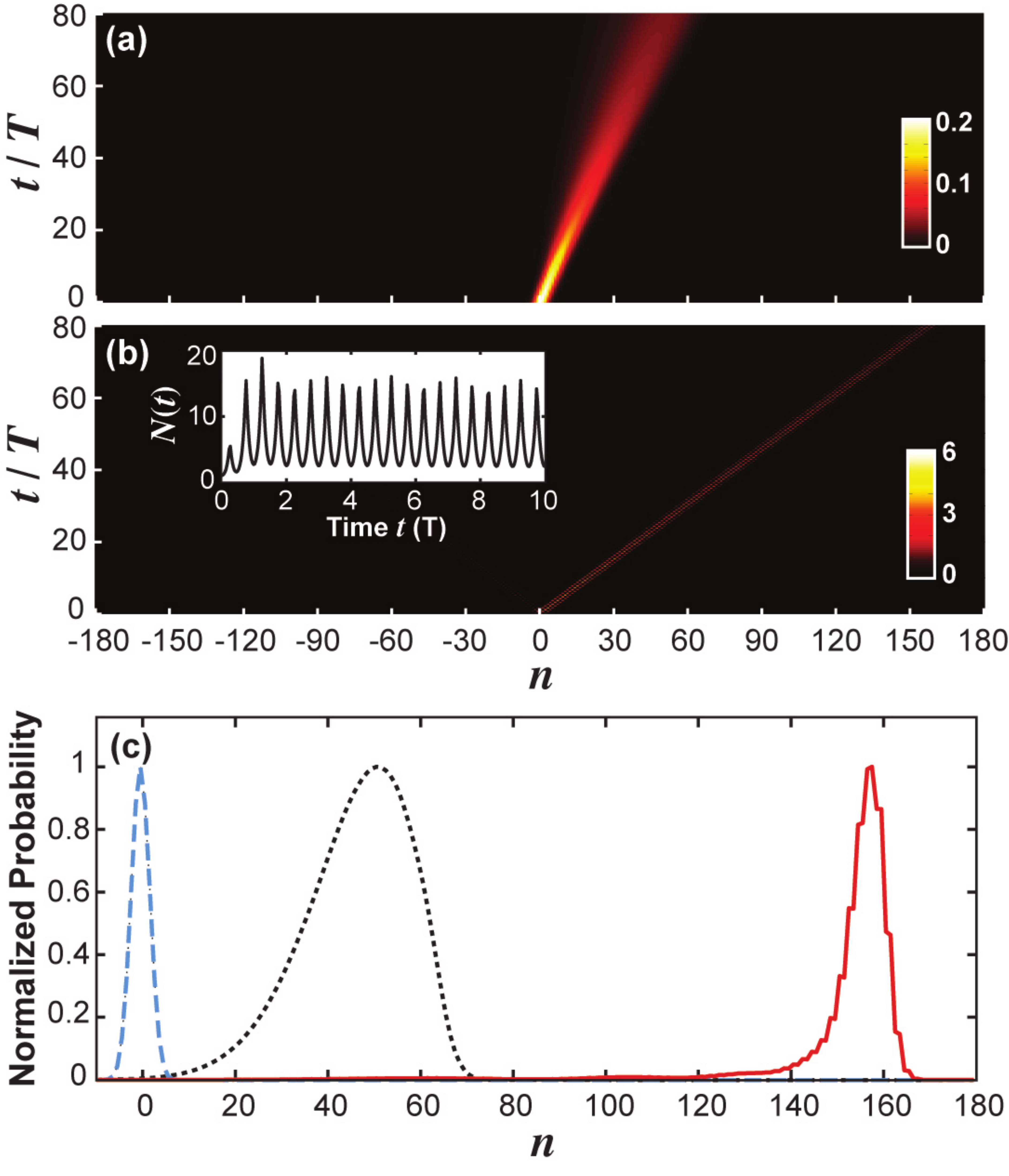}
\caption{(Color online). Occupation probability $|c_n(t)|^2$ for modulation frequency $\omega = 15 \kappa$ and initial excitation with a Gaussian wavepacket $c_n(0) = \exp[-(n/4)^2+i n \pi/4]$ for (a) the static Hamiltonian ($\Delta_0 = 0$) and (b) the driven Hamiltonian close to the $\mathcal{PT}$-symmetry breaking ($\Delta_0 = 11.950\kappa$). Inset in (b) shows the vector state norm for the first 10 periods of the modulation. (c) Occupation probability $|c_n|^2$, normalized to its peak value, at time $t = 80 T$, corresponding to the two cases detailed in (a) (dotted line) and (b) (solid line). The input wavepacket (dashed line) is also shown for comparison.}
\label{unspread}
\end{figure}

Finally, we checked the strong reduction of wave packet spreading due to quantum diffraction resulting from the linearization of the quasi-energy spectrum close to the $\mathcal{PT}$-symmetry breaking. To this aim, we numerically solved Eqs.~(\ref{CMEs}) under initial excitation with a Gaussian probability distribution of full-width at $1/e$ of about $6$ sites and carrier momentum $q_0 = \pi/4$. The results for a lattice with $361$ sites are reported in Fig.~5(a) for the static Hamiltonian and in Fig.~5(b) for the dynamic Hamiltonian with a modulation frequency $\omega = 15 \kappa$ and $\Delta_0 = 11.950$, that is close to the $\mathcal{PT}$-symmetry breaking. Normalized probability distributions at time $t = 80 T$ are also shown in Fig.~5(c), aside with the initial wavepacket distribution (dashed line) for comparison. The spreading of the wavepacket turned out to be dramatically reduced [solid line in Fig.~5(c)] as compared to the spreading in the static lattice [dotted line in Fig.~5(c)].

\section{Conclusions}

We reported on a theoretical investigation of the spectral features and transport properties of a one-dimensional tight-binding superlattice under a time-periodic $\mathcal{PT}$-symmetric modulation of the complex site energies. We observed a $\mathcal{PT}$-symmetry phase transition of the system above a minimum value of the modulation frequency, provided that the amplitude of the modulation reaches a certain threshold value. The threshold turns out to increase almost linearly with the modulation frequency. Interestingly, the unbroken $\mathcal{PT}$-symmetric phase of the driven system exhibits novel spectral features and transport behavior, with  no counterpart in Hermitian time-periodic lattices. In particular, as the $\mathcal{PT}$-symmetry breaking is approached, the quasi-energy spectrum is progressively broadened and linearized. As opposed to the ac-driven control scheme of Hermitian lattices, one of the main property of the complex ac modulation of the $\mathcal{PT}$ lattice is to {\it enhance} (rather than reduce or even suppress) the ballistic motion of the particle, giving rise to an {\it hyper-ballistic regime}, where the speed of propagation in the lattice in not limited by the hopping rate of the static lattice. Moreover, a nearly non-dispersive propagation regime is observed, as confirmed by direct numerical simulations. 

We believe that our results can disclose novel opportunities for coherent control of transport in non-Hermitian lattices by exploiting the unique features of $\mathcal{PT}$-symmetric Hamiltonians in the unbroken phase. In particular, it is envisioned that our approach can find a physical implementation in active optical systems based on synthetic photonic lattices or coupled fiber loops.

\acknowledgments

This work was supported by the Fondazione Cariplo (Grant No. 2011-0338).
\appendix
\section{Quasi-energy spectrum and the monodromy matrix}
In this appendix we provide analytical derivation of the monodromy matrix for the system of Eqs.~(\ref{CMEs}), leading to the quasi-energies of the time-periodic $\mathcal{PT}$-symmetric superlattice. Let us first introduce two new complex variables, for the even-site and odd-site Wannier basis elements respectively:
\begin{subequations}
\begin{align}
& a_n = c_{2n},\\
& b_n = c_{2n+1}.
\end{align}
\end{subequations}
Eqs.~(\ref{CMEs}) are thus split into two sets of coupled equations:
\begin{subequations}\label{CMEsab}
\begin{align}
& i \dot{a_n} = -\kappa \left( b_{n} + b_{n-1} \right) + i\Delta(t) a_n,\\
& i \dot{b_n} = -\kappa \left( a_{n+1} + a_{n} \right) - i\Delta(t) b_n.
\end{align}
\end{subequations}
Since Wannier basis and Bloch basis are related by Fourier transformation, we can expand the Wannier coefficients $a_n(t)$ and $b_n(t)$ in terms of the Bloch spectrum of the vector state defined by $A(q,t)$, $B(q,t)$ with $-\pi \leqslant q \leqslant \pi$, i.e.:
\begin{subequations}\label{ABnq}
\begin{align}
&a_n (t) = \frac{1}{2\pi} \int_0^{2\pi} A(q,t) e^{-iqn} dq,\\
&b_n (t) = \frac{1}{2\pi} \int_0^{2\pi} B(q,t) e^{-iqn} dq.
\end{align}
\end{subequations}
Substitution of Eqs.~(\ref{ABnq}) into Eqs.~(\ref{CMEsab}) yields the following set of coupled equations for the evolution  of the Bloch spectral functions $A(q,t)$ and $B(q,t)$:
\begin{subequations}\label{CMEsAB}
\begin{align}
& i \dot{A} = -\kappa \left( 1 + e^{iq} \right) B + i\Delta(t) A,\\
& i \dot{B} = -\kappa \left( 1 + e^{-iq} \right) A - i\Delta(t) B.
\end{align}
\end{subequations}
When $\Delta(t)$ is periodic of period $T$, for any given value of the quasi-momentum $q$, Eqs.~(\ref{CMEsAB}) represent a homogeneous system of ordinary differential equations with time-periodic coefficients. According to Floquet theory, the solution of this system has the general form (see for instance \cite{Teschl_book}):
\begin{equation}
\left[\begin{array}{c}
A(q,t) \\
B(q,t)
\end{array}\right] = {\bf P}(q,t-t_0) e^{i (t-t_0) {\bf R}(q)} \left[\begin{array}{c}
A(q,t_0) \\
B(q,t_0)
\end{array}\right],
\end{equation}
\noindent where $[A(q,t_0) \; B(q,t_0)]^T$ is the initial value at time $t_0$, ${\bf P}(q,t-t_0)$ is a $2\times2$ time-periodic matrix of period $T$ with ${\bf P}(q,0)$ being the identity matrix, and ${\bf R}(q)$ is a time-independent matrix whose eigenvalues are the quasi-energies (so-called Floquet exponents) of the system. The real and imaginary parts of the quasi-energies can be computed from the eigenvalues $\eta(q)$ of the monodromy matrix ${\bf M}(q)$  using Eqs.(7) given in the text. The monodromy matrix connects the solution to Eqs.~(A4) over one oscillation cycle (i.e. from $t=t_0$ to $t=t_0+T=t_1$) and is given by ${\bf M} = e^{i T {\bf R}}$. The columns of ${\bf M}$ are the two solutions $[A_1(t_1) \; B_1(t_1)]^T$ and $[A_2(t_1) \; B_2(t_1)]^T$ of Eqs.~(A4) obtained under initial conditions (at time $t_0$) $[1 \; 0]^T$ and $[0 \; 1]^T$ respectively. Since for a square wave modulation the coefficients in Eqs.~(\ref{CMEsAB}) are constants over a semi-cycle, $A_{1,2}(t_1)$ and $B_{1,2}(t_1)$ can be analytically determined over each semi-cyle, thus ${\bf M}$ factorizes into the product of the two matrices (one for each semi-cycle) ${\bf M}_1$ and ${\bf M}_2$ given by Eqs.~(\ref{M1M2}).

\end{document}